# Fixed-Mobile Convergence in the 5G era: From Hybrid Access to Converged Core

Massimo Condoluci, *Member, IEEE*, Stephen H. Johnson, Vicknesan Ayadurai, Maria A. Lema, Maria A. Cuevas, Mischa Dohler, *Fellow, IEEE,* and Toktam Mahmoodi, *Senior Member, IEEE*

*Abstract—* **The availability of different paths to communicate to a user or device introduces several benefits, from boosting end-user performance to improving network utilization. Hybrid access is a first step in enabling convergence of mobile and fixed networks, however, despite traffic optimization, this approach is limited as fixed and mobile are still two separate core networks inter-connected through an aggregation point. On the road to 5G networks, the design trend is moving towards an aggregated network, where different access technologies share a common anchor point in the core. This enables further network optimization in addition to hybrid access, examples are user-specific policies for aggregation and improved traffic balancing across different accesses according to user, network, and service context. This paper aims to discuss the ongoing work around hybrid access and network convergence by Broadband Forum and 3GPP. We present some testbed results on hybrid access and analyze some primary performance indicators such as achievable data rates, link utilization for aggregated traffic and session setup latency. We finally discuss the future directions for network convergence to enable future scenarios with enhanced configuration capabilities for fixed and mobile convergence.**

## I. INTRODUCTION

T HE trend of ever increasing availability of heterogeneous accesses offers an opportunity for exploitation in order to boost network capacity and present new business opportunities [1]. The possibility of simultaneously using fixed and mobile broadband (FBB and MBB, respectively) introduces benefits, first of all in terms of boosting end-user performance but also in terms of load balancing, network optimization, "always best connected" and session continuity when leaving coverage of one access, i.e., failover. The exploitation of multiple accesses can be accomplished via different approaches to achieve different benefits. A first option is to exploit the aggregation of multiple available links towards one destination to boost the data rate of a session. As depicted in Fig. 1(a), this can be achieved by managing the available links at the end-points (i.e., the end-device and the remote server) thanks to utilization of transport protocols such as multipath TCP (MPTCP) [2] for advertising the availability of multiple links and managing traffic aggregation. In this case, the end-points are unaware of the status of the two networks, and the FBB and MBB links are unaware of the aggregation that is performed. Therefore, this limits the opportunities for network operators to optimize the traffic within their networks.

In order to exploit in a more effective way the availability of multiple links, the attention has moved towards the so-called *fixed-mobile convergence* (FMC), where a service provider is aware of the availability of the different links and can exploit this capability directly in a coordinated way. FMC enables the possibility to deliver any service, anywhere and via any access technology. Given the fact that, historically, FBB and MBB networks have been built separately and operated independently, FMC can be realized by exploiting a hybrid access gateway (HAG) [3]. As depicted in Fig. 1(b), the HAG acts as an aggregation point for the traffic from/to the Border Network Gateway (BNG) and the Packet Gateway (PGW) of the fixed and mobile networks, respectively. In this case, an operator managing the two networks can be aware of the availability of multiple accesses and, thus, properly manage the traffic across them [4]. In addition to improved user performance, the exploitation of a HAG has benefits in terms of: (i) improved reliability, as traffic can be switched to another access if the performance degrades due to either congestion or mobility issues; (ii) seamless experience, as a service can now be accessed via either FBB or MBB. The latter aspects might also enable rapid service deployment by providing services such as high-speed broadband usually delivered via, e.g., fixed access through mobile networks in regions where fixed access is difficult and costly (or vice versa). Nevertheless, some limitations are due to the fact that the BNG/PGW nodes have a limited visibility of the network (e.g., layer 1/2 and mobility status is not available) and this limits the possibilities for the overall optimization of the end-to-end paths.

Recently, the 3rd Generation Partnership Project (3GPP) has been working towards the concept of having a *converged core network* on the road to the standardization of the 5G system architecture [5]. With this solution, shown in Fig. 1(c), one core network (CN) simultaneously manages different

M. Condoluci, M. A. Lema, M. Dohler, and T. Mahmoodi are with the Centre for Telecommunications Research, Department of Informatics, King's College London, UK. E-mail: {massimo.condoluci, maria.lema_rosas, mischa.dohler, toktam.mahmoodi}@kcl.ac.uk.

V. Ayadurai is with Ericsson Research, Stockholm, Sweden. E-mail: vicknesan.ayadurai@ericsson.com.

S. Johnson and M. A. Cuevas are with British Telecommunications plc, UK. E-mail: {stephen.h.johnson, maria.a.cuevas}@bt.com.

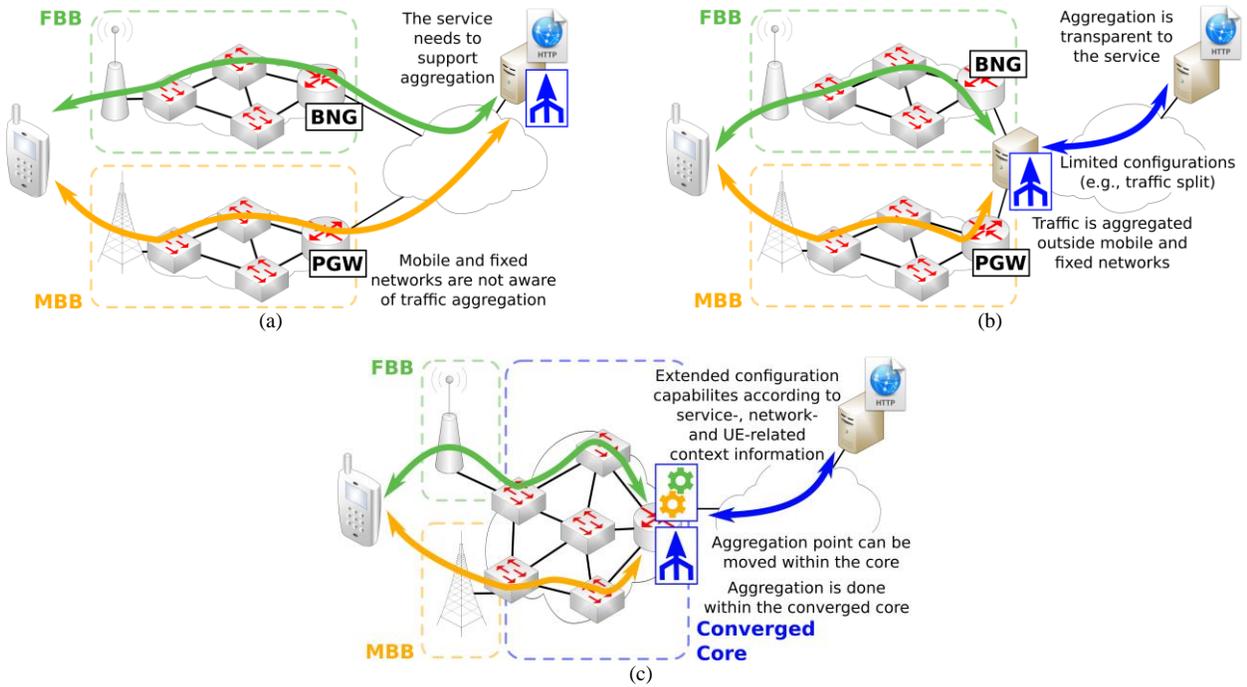

Fig. 1. Exploiting FBB and MBB from an architecture point of view: (a) traffic aggregation performed at the end-side service, (b) FMC with a HAG inter-connecting fixed and mobile networks, (c) FMC with converged core network.

radio access technologies instead of having two separated CNs managed by a HAG. This capability advances the features of FMC by enabling more accurate access traffic steering, switching, and splitting (ATSSS) [6] policies by exploiting service-, network-, and user-related information available at the CN. This feature allows the definition of policies to map specific services to the access(es) that better support their QoS requirements. This is of particular importance to effectively deliver services with strict QoS constraints, e.g., ultra-reliable low latency communications (URLLC), as it avoids delivering the service via an access that cannot support it. This allows service providers to apply the best connectivity for optimal network utilization in addition to enhancing end-user experience. Another advantage is that the network operator can guarantee consistent performance across the user equipment (UE) as all UEs equipped with the same protocol stack will be able to support the same set of ATSSS policies.

The aim of this paper is to present the status of FMC activities, focusing on both hybrid access and converged CN approaches in order to discuss their capabilities in terms of network configuration. We will summarize the state of the art for hybrid access as well as discussing the recent advances by the 3GPP in supporting the enforcement of ATSSS policies in converged networks. We will also analyze the role played by MPTCP in hybrid access and converged core scenarios as well as the improvements needed in order to better cope with these approaches. We have developed a testbed where a HAG aggregates the traffic of a FBB and an in-house commercial-grade LTE network on which we conducted some experiments to understand how capacity and latency imbalances between FBB and MBB links affect the overall performance when using MPTCP to aggregate the traffic. We will also analyze the benefits of FMC in reducing the session setup time, as this aspect has some benefits in improving the network utilization and the radio link efficiency. We will finally discuss the future directions to be investigated in the design of ATSSS capabilities within the converged 5G core architecture.

## II. HYBRID ACCESS: A SURVEY AT A GLANCE

### A. BBF View

The Broadband Forum (BBF) is currently investigating different aspects related to hybrid access with the aim to define and examine scenarios, evaluate possible business models, and drive the technical roadmap to align the industry and to interwork with other standardization bodies.

The reference architecture to move from non-converged networks towards a fully hybrid access network is presented in [3]. As depicted in Fig. 1(b) and in more details in Fig. 2, coordinated and simultaneous access over different networks is offered to a hybrid customer premises equipment (HCPE) through a hybrid access gateway (HAG). The HCPE aggregates and distributes traffic in the upstream while the HAG, which can be placed either at the BNG or PGW, performs aggregation and distribution mechanisms for downstream traffic. As depicted in Fig. 2, QoS and policy enforcement for upstream and downstream can be implemented at the HCPE or the HAG, respectively.

To carry the traffic between the HAG and the HCPE, BBF mainly considers different transport options: layer 3 (L3) overlay tunneling, L3 network-based tunneling and L4 multipath, like MPTCP.

### B. IETF View

The Internet Engineering Task Force (IETF) has been focusing on supporting multiple paths from a transport protocol point to view. The IETF Transport Area designed

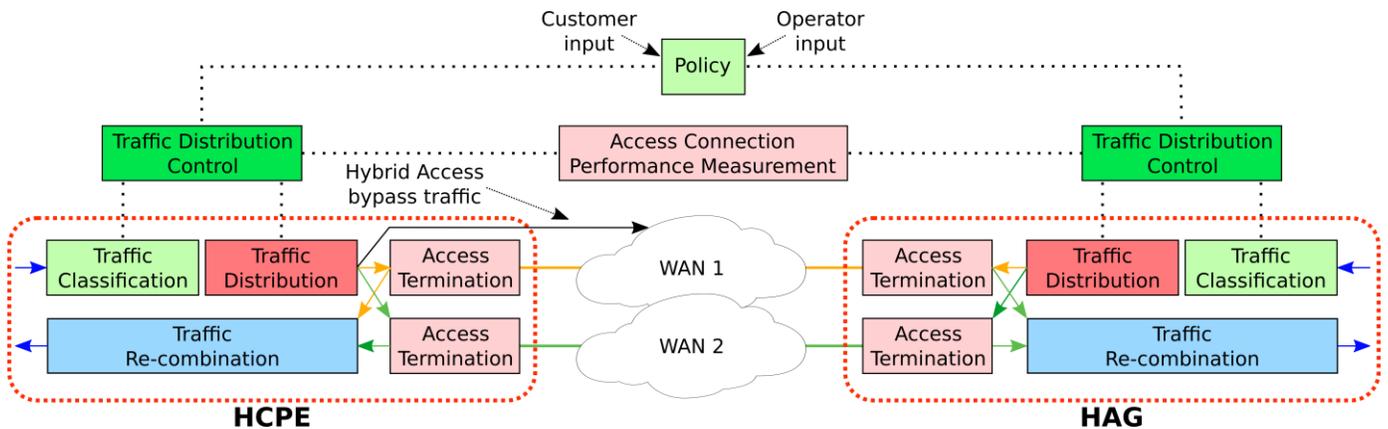

Fig. 2. Hybrid Access reference architecture from BBF [3]. This architecture highlights the role of the HAG in inter-connecting different wide area networks (WANs).

MPTCP for the twofold purpose of improved throughput and resiliency [2]. By building upon widely used TCP and having stringent networking compatibility goals, MPTCP integrates well into many networks and coexists easily with middleboxes like firewall and Network Address Translation (NAT) already in place, which helped to expedite its uptake.

The MPTCP working group documents "real world" experiences of MPTCP playing many diverse roles over the years such as load-balancing in datacenters, FBB/MBB offloading and as MPTCP proxies [7]. These last two capabilities enable MPTCP to play a prominent role in the area of hybrid access.

The primary goal of the MPTCP working group is to create a bis version of the protocol document on the Standards track, approximately by the end of Q1 2018. At the time of writing, current MPTCP specifications are defined in [8].

### C. State of the art

FMC has been an area of interest for network operators for some time. One of the major research efforts on the FMC subject has been pursued in the context of COMBO project[1], which has studied both structural and functional aspect of the convergence with the assumption that FBB and MBB networks are separated. However, in the 5G era, alternative means of hybrid access are being investigated to satisfy the different services. Work in [4] describes in detail the hybrid access use cases and provides a comprehensive overview of the different transport options discussed in both 3GPP and BBF standardization bodies. Authors in [4] focus on the architectural concepts and review the implications in control- and user-plane (CP and UP, respectively). More specifically, the focus is on how policy and charging mechanisms defined by 3GPP and BBF can inter-work. In addition, [4] discusses the role of programmable networks enabled by Network Functions Virtualization (NFV) in supporting enhanced traffic distribution schemes.

Similarly, authors in [9] explain the 3GPP and BBF interworking architecture using WLAN access or femto-cells and focus on the way towards a unified network for policy and

QoS convergence, however authors do not mention the hybrid access use case of network convergence.

Probably the most relevant work on hybrid access via MPTCP is the one presented in [10]. Authors discuss the benefits in terms of bandwidth aggregation, increased reliability and service continuity, all facilitated by implementing a carrier-grade MPTCP proxy which allows service providers to implement traffic steering policies using the available access networks.

### III. 3GPP EFFORTS ON NETWORK CONVERGENCE

3GPP is currently investigating how to simultaneously inter-connect both 3GPP and non-3GPP accesses to the 5G core. In addition, 3GPP recently started focusing on designing an architecture and related procedures to take advantage of the availability of multiple accesses.

### A. Support of non-3GPP access

From an architecture point of view, the support of untrusted non-3GPP access is discussed in [5] and it is depicted in Fig. 3. 3GPP introduced a Non-3GPP InterWorking Function (N3IWF) in Release 15, providing CP/UP functions for untrusted non-3GPP accesses[2]. In the CP, the N2 interface connects the N3IWF to the Access and Mobility Management Function (AMF, in charge of registration, connection, reachability, mobility, etc.). The N3 interface connects the N3IWF to the User Plane Function (UPF, in charge of packet routing and forwarding, traffic usage reporting, etc.), which will thus represent the convergence point in the core of both 3GPP and non-3GPP accesses described earlier in Fig.1(c). Finally, the N3IWF allows devices accessing the 5G core through non-3GPP access to support non-access stratum (NAS) signaling through the N1 interface.

The availability of a CP connection via the N3IWF allows the CN to manage a UE connected via a non-3GPP access in a similar way as it were connected via 3GPP access thanks to the availability of functionalities such as NAS signaling, address allocation, policy enforcement, etc. This effectively simplifies the management multiple accesses in a coordinated way. Indeed, having 3GPP and non-3GPP accesses anchored to the same CN means that traffic, both within the core as well as within the radio access network can be optimized in a more effective way on an end-to-end basis. The UPF now has an overall vision of the traffic from/to all anchored accesses while the AMF will have visibility of the status of the available radio links. Actions such as load balancing and access selection can be performed with more accurate information thus optimizing network utilization and end-user performance.

---

[2] Support for trusted non-3GPP access in the 5G system architecture is being currently discussed in TR 23.716 and will be included in Release 16.

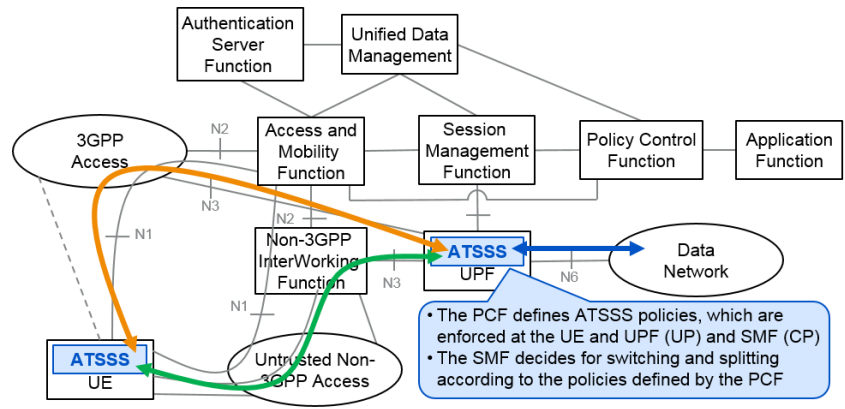

Fig. 3. Architecture of 5G core network with non-3GPP access [5], enhanced with ATSSS capabilities [6].

### B. Access traffic steering, switching and splitting (ATSSS)

3GPP is investigating how to enable the management of traffic across multiple accesses connected to the same core. To this aim, three traffic functions have been identified [6]: *steering*, i.e., the selection of the access technology to steer a network/UE-initiated traffic; *switching*, i.e., moving an ongoing flow from one access to another; *splitting*, i.e., the simultaneous use of different accesses (i.e., aggregation).

The access traffic steering, switching and splitting (ATSSS) between 3GPP and non-3GPP accesses is depicted in Fig. 3, where ATSSS functionalities reside on both CN and UE sides. From the CN point of view, the ATSSS functionalities are:

- *Serving as anchor point for flows that can be potentially switched or split.*
- *Defining policies for ATSSS.* Integrated policies for both FBB and MBB can be designed given the fact that the same core network manages both accesses. Policies are provided by the Policy Control Function (PCF) and might depend on the type of service a flow belongs to, UE's subscription and context information, impact on other ongoing traffic, etc. ATSSS policies may be defined at the service level and may include a prioritized list of accesses as well as criteria necessary to drive switching or splitting according to, for example, signal strength, UE context-related (speed, location, etc.) or performance metrics (throughput, latency, packet losses, jitter, etc.).
- *Conveying ATSSS policies to the UE.*
- *Monitoring ongoing flows.* This is performed by the

UPF where the parameters to monitor could be pre-defined or policy-based.

- *Session management.* The SMF is in charge of managing ongoing ATSSS sessions and taking decisions to either switch or split an ongoing session in response to network changes (e.g., link failure, congestion).

The SMF is the CP enforcement point of ATSSS policies. The UE and UPF are the UP enforcement points for UE- and network-initiated data flows, respectively. The definition of the architecture for ATSSS and related procedures is still an ongoing process, where recent updates and a list of key issues can be found in [6].

### C. Multi-path transport protocol for 3GPP ATSSS

In the 3GPP ATSSS framework, the CN is in charge of the management of the available links such as selection of the access(es) to be used, access advertising, etc. Nevertheless, the exploitation of a transport protocol able to simultaneously manage multiple links might introduce several benefits in allowing a quick switching/splitting capability while guaranteeing session continuity.

The effective implementation of ATSSS features requires the ATSSS functionalities to interact/instruct the transport protocol. For TCP flows, MPTCP represents a possible solution but several enhancements could be introduced to further boost the performance and network exploitation capabilities. A first aspect to consider is to ensure that the CN/ATSSS is aware of the IP address/access relationship, information that is not available with the MPTCP address advertisement mechanism. In addition, the exploitation of a 3GPP protocol stack at the UE/CN means that common packet filter rules can be used to enforce service-oriented policies, thus enhancing the flexibility of MPTCP in providing service-specific treatment of the traffic. Another aspect to consider is related to the address advertisement mechanism, which allows MPTCP to discover the available paths towards the end-point destination. In case of a converged CN with ATSSS capabilities, the network is already aware of the available accesses for a given UE (each with a different address) and this information can be exploited by all sessions originated by the UE. This means that, in case of a converged CN with ATSSS, some of the messages for address advertising might be not needed, thus allowing ATSSS traffic to start immediately via multiple paths.

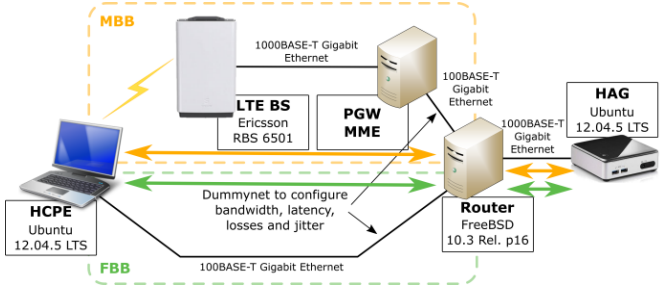

Fig. 4. Architecture of our testbed platform.

To enable the above discussed features, ATSSS entities at the UE and the UPF should maintain an API with the transport protocol used to carry UP traffic to optimize and drive its behavior. To this end, MPTCP provides a set of APIs [11] to, e.g., indicate the interface to be used as a default path and send commands to change the priorities of the available sub-flows.

## IV. TESTBED: HYBRID ACCESS CAPABILITIES

The aim of our testbed, depicted in Fig. 4, is to create a platform enabling hybrid access over FBB and MBB. The testbed comprises of physical LTE, 4G core network and Ethernet links, that provide a realistic setup. In order to conduct an analysis considering performance as in real deployments, our testbed has been configured by taking into account statistical analysis FBB and 4G MBB performance in UK provided by Ofcom [12], [13].

### A. Testbed description

The testbed consists of an MPTCP-capable HCPE which communicates with an MPTCP-capable HAG via an operator which runs FBB and MBB. HCPE and HAG are Linux-based computers with an experimental in-house developed implementation of MPTCP. A FreeBSD-based router acts as a common anchor point for FBB and MBB and connects both networks to the HAG via a high-speed Gigabit Ethernet to accommodate traffic from/to both networks.

The HCPE is multi-homed with MBB and FBB interfaces. The MBB link is made via an LTE dongle which dials up to an in-house commercial-grade LTE eNB. Traffic from/to the eNB is managed by a PGW for the UP traffic and by a Mobility Management Entity (MME) for the control-plane traffic. MBB's RTT is ~53ms[3], with throughput performance ~20Mbps and ~5Mbps in downlink and uplink, respectively. These values are compliant with 4G performance in UK [12].

The second interface of the HCPE is connected via Ethernet directly to the FreeBSD router. Employing the *dummynet* tool on this interface, we emulate the FBB environment denoted as "up to 76Mbps FTTC" in [13], with a RTT equal to ~13ms and average rates as shown in Fig. 5(b).

The HCPE and the HAG act as the end-points in our

experiments, hosting the applications used for testing. TCP-based traffic[4] is managed by running several applications targeting the transmission of a 100MB file: *scp* has been used in both downlink and uplink directions to consider an application with a complex session setup mechanism; *iperf* has been used for HCPE-HAG traffic; *wget* for HAG-HCPE traffic. Network analysis tools such as *tcpdump* and *ifstat* are used to carry out measurements over the system.

### B. Link utilization analysis

Fig. 5 focuses on the link rate for the cases without FMC (i.e., MBB only and FBB only) and with FMC (i.e., FBB and MBB used simultaneously). Fig. 5(a) provides an example of the link rate for a scp session in the downlink. For the FMC case, in Fig. 5(a) the top plot depicts the aggregated link rate while the plot on the bottom shows the link rates of FBB and MBB components of the FMC session. Although this example shows an expected result, i.e., FMC allows a quicker file download thanks to the simultaneous use of FBB and MBB links, it also shows that the link rate is less stable compared to the case when FBB and MBB are used without FMC. This behavior is due to fact that MPTCP has to manage the traffic of two links with different capacities and RTTs, and this might have the side effect of reducing the utilization of one link if an acknowledgment from another path is missing due to losses or delays (as in [14], the congestion window of one subflow depends on the acknowledgment of the whole MPTCP connection, i.e., of all subflows). In fact, as seen on the bottom of Fig. 5(a), MPTCP reduces the rate of the traffic on the FBB link due to the capacity/RTT imbalance compared to the MBB link. This aspect is further analyzed with different applications in Fig. 5(b), which shows the link rate of FBB and MBB when they are used alone (left) and when they are aggregated in case of FMC (right). This analysis shows that, when having FMC, the downlink and uplink link rates are ~80Mbps and ~18Mpbs, respectively, meaning that the aggregated link rate is not equal to the sum of the link rates of aggregated accesses. Comparing the performance of FBB and MBB when aggregated with FMC, the utilization of the FBB and MBB links is reduced down to 85% and 82%, respectively, compared to the case when FBB and MBB are used without FMC. The impact is, on average, more remarked in downlink which is also the direction with the highest imbalance when comparing the capacity and the latency values of the two accesses. Another aspect to consider is that while FBB's behavior in case of FMC does not meaningfully change in case of different applications, MBB's links are less utilized in case of wget and iperf w.r.t. the scp case. Summarizing above results, our testbed shows that FMC effectively improves UE's performance in terms of data rate, but also the imbalances in capacity/RTT of the available links affect the gains in terms of aggregated performance and this depends on the transport protocol managing the traffic aggregation.

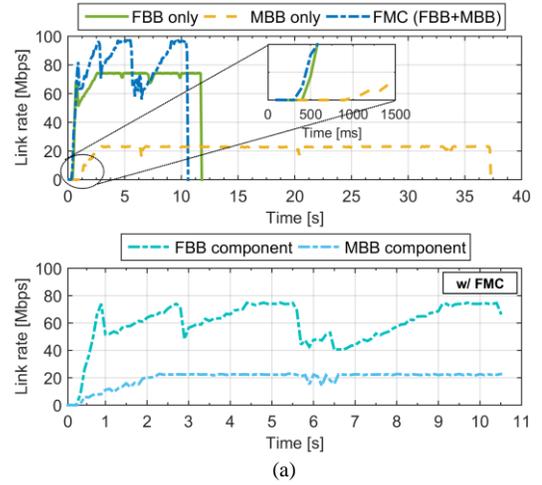

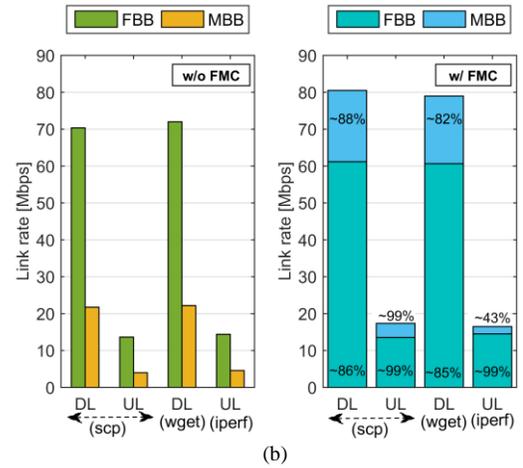

Fig. 5. Link utilization: (a) example of scp in downlink, (b) average results for different directions and applications.

[4] The congestion algorithm for each subflow is CUBIC. The congestion control at the MPTCP connection level is based on [14], where the aim is to improve throughput and balance congestion among the subflows. This is achieved by jointly setting the increase of the congestion window of each subflow according to the MPTCP connection level acknowledgement.

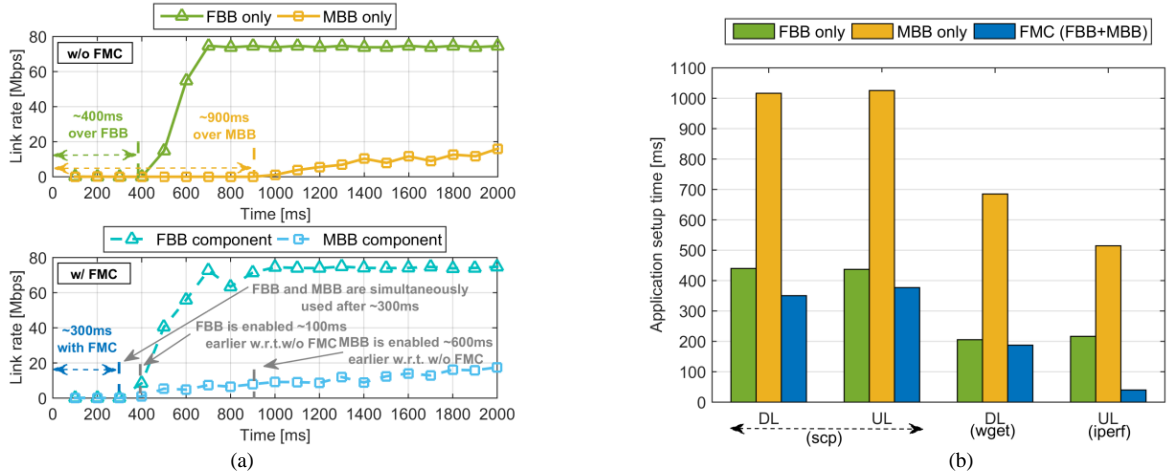

Fig. 6. Application setup time: (a) example of scp in downlink, (b) average results for different directions and applications.

One aspect to be further highlighted from the results in Fig. 5 is the initialization phase of the sessions. From Fig. 5(a), we can observe that the effective application traffic (i.e., after the initial phase of setting up the connectivity and establishing the session from an application point of view) starts earlier in the case of FMC than for the cases of FBB and MBB only. This aspect will be further analyzed in the following.

### C. Session setup analysis

One aspect analyzed in our testbed is the session setup time, measured as the time interval from the moment the application is triggered to the moment the first data packet is transmitted over the *radio* interface. This analysis allows to understand two aspects: *(i)* the "waiting" time for the end-user before the effective session start; *(ii)* the utilization of the available links. The second aspect is interesting from a network point of view, because session setup is composed of small packets for, e.g., key exchange, mutual authentication, and this might impact the overall utilization of the links and especially the spectral efficiency of MBB.

From the analysis for scp in the downlink case, the top plot in Fig. 6(a) shows that the session setup time for FBB and MBB only cases are ~400ms and ~900ms, respectively. The bottom plot in Fig. 6(a) shows that in case of FMC the application traffic starts after ~300ms, i.e., both FBB and MBB links can be used ~100ms and ~600ms earlier compared to the case without FMC. This behavior is given by the fact that also the session setup can happen in parallel via the different links with FMC, i.e., FMC further improves the user performance such that the user will be able to start a session earlier than if having a single FBB or MBB link. In addition, from a network point of view, FMC allows resources to be used more efficiently for the transmission of (possibly huge) data packets instead of being underutilized for a long-time interval.

Fig. 6(b) shows the average sessions setup time for different applications, which testifies that FMC achieves a session setup time reduction also in case of sessions with lighter setup procedures (i.e., wget and iperf) compared to scp.

### V. INSIGHTS AND FUTURE DIRECTIONS

From the results presented in the previous Section, we can conclude that:
- FMC allows boosting of UE performance, but the strategy exploited to simultaneously utilize multiple

links might limit the gains in terms of aggregated data rate if unbalanced capacity/RTT figures are not managed properly.

- FMC can exploit the availability of different accesses to shorten the session setup time, with consequent benefits from an end-user and network efficiency points of view.

The following aspects need to be considered for future improvements.

### A. Link with policy framework and enforcement

As discussed above, policies to drive the access selection for a new session or to change the access for an ongoing session might be useful to allow operators to properly optimize the traffic within their networks as well as to guarantee that a given session is served via the most suitable access according to service requirements and network status.

Supporting policies for ATSSS requires several enhancements. Firstly, policies for FMC need to be integrated within the existing policy framework, in order to allow policies to be enforced in both CP and UP (i.e., SMF and UE/UPF, respectively). To generate subscriber-oriented policies, the policy framework needs to take into consideration information such as subscriber priority, to guarantee that a given subscriber will be served according to their profile. In addition, to effectively satisfy session requirements, the policy framework needs to retrieve application-oriented policy (e.g., minimum requested data rate, maximum tolerated latency) to avoid steering or switching a session to use an access that is unable to satisfy the service needs. This aspect is of particular interest in order to effectively support services with strict QoS requirements such as URLLC. This also means that information about the ability to satisfy some traffic requirements needs to be available at the network side.

Once policies are generated and enforced, switching and splitting can be used to react to changes such as mobility, access availability and congestion. To enable this capability, two components are needed: *(i)* monitoring of ongoing sessions to allow the network to understand if service requirements are currently satisfied by the selected access; *(ii)* link with UP policy enforcement points (i.e., UE and UPF) to enable switching/splitting of existing traffic. These two aspects need to be properly investigated, especially regarding the delivery of policies and commands to the UE from the network.

### B. Management of multiple links

As shown in our experiments, the presence of links with unbalanced capacity/RTT might result in underutilization of available accesses compared to the cases when such accesses are managed without aggregation (i.e., no FMC). Nevertheless, even when the links are balanced in terms of average capacity and delays, issues such as multi-user multi-link scheduling becomes very relevant in large scale deployments. Although FMC should be independent of the transport protocol used to aggregate the available links, effective solutions to avoid the reduction of the link utilization when aggregating multiple links are thus beneficial to further boost the performance of FMC. An example on this direction can be found in [15], which investigates the problem of load distribution for real-time traffic over multipath networks and proposes a goodput-aware load distribution model that takes into consideration path status estimation to accurately sense the quality of each transport link, flow rate assignment to optimize the aggregate goodput of input traffic, and deadline-constrained packet interleaving to mitigate consecutive losses. Another aspect requiring further investigation is related to the size of data packets, as MPTCP-like solutions might be better designed for large packets while other approaches with ideally less overhead should be considered for traffic dealing with small-packets, as for instance analyzed in [2] and [16], where the latter considers low-latency aspects as well. Nevertheless, an effective multi-link exploitation to increase reliability and achieve low-latency for URLLC should be further investigated.

In addition, FMC should work regardless of the transport protocol type. Currently, MPTCP is considered as a possible solution for TCP connections, while further improvements are currently under investigation for UDP. Other approaches, such as QUIC, might offer alternative solutions to manage multiple links for UDP connections with less overhead compared to MPTCP.

## VI. Conclusions

This paper discussed the ongoing activities on convergence between fixed and mobile networks. Different solutions have been presented considering the standardization efforts by BBF and 3GPP, with focus on hybrid access and aggregated core network with ATSSS capabilities. On the latter aspect, the paper described the further benefits introduced with ATSSS in terms of enforcing policies for traffic management in a unified way in both FBB and MBB. Future directions have been identified and discussed, which can be used to drive the research on this topic.

A testbed has been developed to analyze some preliminary results achieved with FMC, focusing on aspects such as link utilization and session setup time when using MPTCP as a transport protocol to aggregate FBB and MBB. In addition to boosting end-user performance in terms of throughput, results underlined the benefits introduced with FMC in reducing the session setup time, with consequent gains in enhancing network utilization and radio link efficiency. Further improvements can be achieved by introducing an enhanced congestion control mechanism to cope with links that have unbalanced capacity and RTT values.

## Acknowledgments

This work has been supported by the industry grant to King's College London "Low latency Research: Beyond Convergence" funded by British Telecommunications plc and by the industry grant to King's College London "5G and Tactile Internet" funded by Ericsson.